\documentstyle[12pt]{article}

\begin{document}

\title{Evolution of DIS structure functions at small $x$}

\author{B.I. Ermolaev\\
{\small A.F. Ioffe Physico-Technical Institute, 194021 St.Petersburg, Russia}
}
\date{}
\maketitle

In the present talk I discuss the recent progress in the theoretical
description of the Deep Inelastic Scattering (DIS) at small $x$. 
From the theoretical point
of view, the inclusive cross-section of DIS can be regarded as a product of
the leptonic and the hadronic tensors, with exchange by a deeply off-shell
virtual photon. Further on I discuss only the hadronic
tensor. The hadronic tensor consists of two parts: the spin-dependent part and
the spin-independent one. My talk is about the spin-independent part,
$W_{\mu\nu}$. This tensor should be made of on-shell momenta $P$
(the initial/final hadron momentum) and of the deeply off-shell momentum
$q$ (the incoming/outgoing virtual
photon momentum). Also it should bear the Lorentz indices $\mu$ and $\nu$
corresponding to the virtual photon polarizations. It must also respect
both the gauge and the Lorentz invariance. This leaves us with two independent
projection operators, each multiplied by a scalar function. These are the
spin-independent structure functions $F^h_1$ and $F^h_2$. Their arguments
must be some
scalars made of $P$ and $q$. Traditionally, they are chosen as
$Q^2 \equiv -q^2 >0$ and $X = Q^2/2Pq$, $(0<X<1)$. Thus,

\begin{equation}
\label{wh}
W_{\mu\nu} = (-g_{\mu\nu} + \frac{q_{\mu}q_{\nu}}{q^2}) F^h_1(X, Q^2) +
(P_{\mu} - q_{\mu}\frac{Pq}{q^2})(P_{\nu} - q_{\nu}\frac{Pq}{q^2})
\frac{1}{Pq}F^h_2(X,Q^2) ~.
\end{equation}

The next step is exploiting the factorization.
According to it, the whole hadronic tensor, and its spin-independent part in
particular, can be regarded as a convolution of two objects:
the probability $P_h^p$ to find a parton (a quark and
or a gluon)  and the Deep Inelastic Scattering off the
parton. Such partons are supposed to be nearly on-shell, so one can construct
the "partonic" tensor for them in the same way as Eq. (\ref{wh}) was made,
replacing $P$ by $p$ and $X$ by $x$.
Such new spin-independent tensor $w_{\mu\nu}$ (parameterized by
"partonic" structure
functions $F_1$ and $F_2$, each of them depends on $Q^2$ and $x = Q^2/2pq$),
can be calculated with the perturbative
QCD methods and must be convoluted with $P_h^p$ . As for calculating
$w_{\mu\nu}$, it is a purely technical problem, though quite complicated.
However, it is scarcely possible to convolute $w_{\mu\nu}$ with $P_h^p$  
because presently there is no model-independent way to calculate $P_h^p$. 
Instead, the regular way to obtain $F^h_{1,2}$ is imitating this procedure by
 picking up good fits to fill
in the gap between $W^S_{\mu\nu}$ and $w_{\mu\nu}$.
 From now on I discuss $w_{\mu\nu}$, i.e. DIS off a parton. I choose a quark as
 this parton.
In order to calculate $w_{\mu\nu}$, one have to calculate all involved
Feynman graphs and sum up their contributions. In the Born approximation

\begin{equation}
\label{born}
F_1 = (e^2_q/2)\delta(1 - x),~~~F_2 = 2xF_1 .
\end{equation}
where $e_q$ is the electric charge of the initial quark.\\
The radiative corrections to the Born approximation can be calculated in the
flavour-dependent way where the quark interacting with
the virtual photon is the initial quark
and in the flavour-independent way where these quarks are different. In this
case, there is no dependence on the flavour of the initial quark because
one can sum over flavours of the quark interacting with the virtual photon.
Therefore, any structure function has both
flavour-dependent (non-singlet) and flavour-independent (singlet) parts.
Basically, the singlet parts of the spin-independent structure functions
are more important because they define the
total cross-sections. However, the non-singlet parts are also interesting to
know. For example, one can subtract a structure function for DIS off
proton from the same structure function for DIS off neutron.
The result is flavor-dependent. Then, the non-singlet structure functions
and the singlet ones have some common features. On the other hand,
the non-singlet
structure functions are simpler objects to calculate. So, they can be
regarded as a test field for checking various assumptions
or models made for the singlet structure functions.
The most important radiative corrections to the Born values (\ref{born})
are logarithms  of $x$ and $Q^2$.
When $x$ is large, logarithms of $x$ are not important but when $x$ is
small, none of these kinds of logarithmic contributions must be
neglected a priori.
The less is value of $x$, the more important are contributions of
higher-order radiative
corrections. With certain approximations, there were developed regular methods
to calculate them to all orders in $\alpha_s$.
 The most wide-spread methods among them are the
DGLAP \cite{dglap} and the BFKL \cite{bfkl}. Both of them are
one-dimensional linear differential equations with integral operators in their
right-hand sides. However, they were developed for operating in different
kinematical regions:
the DGLAP was suggested for studying the kinematical region $x\sim 1$ and
large $Q^2$.
It neglects therefore logarithms of $x$ which are not multiplied by
logarithms of $Q^2$ . So, it respects $Q^2$-evolution only.
For evolution of the singlet part of $F_2 \equiv F$ the DGLAP equation can
be written as

\begin{equation}
\label{dglap}
\frac{\partial F}{\partial\ln Q^2} = PF
\end{equation}
where $P$ is the DGLAP integral operator, with the splitting functions as a
kernel.
The splitting functions are known to the $\alpha_s^2$ -order.
After the Mellin transform
\begin{equation}
\label{mellin}
F(x,Q^2) = \int_{-\imath\infty}^{\imath\infty}\frac{dN}{2\pi\imath}
e^{N\ln(1/x)} F(N, Q^2) ,
\end{equation}
the DGLAP is
\begin{equation}
\label{ndglap}
\partial F(N,t)/\partial t = \gamma(N) F(N,t)
\end{equation}
where $t =\ln Q^2$. $\gamma(N)$ are the anomalous dimensions. They are
known to 
the first two orders in  $\alpha_s$.\\
The BFKL was done for accounting for $x$-evolution. It sums up $\ln x$
to all orders in $\alpha_s$. Like the DGLAP, it is also one-dimensional
evolution equation:

\begin{equation}
\label{bfkl}
\frac{\partial F}{\partial\ln (1/x)} = F_0 + KF
\end{equation}
where $K$ is the BFKL integral operator,  and $F_0$ is the singlet $F_2$ in
the lowest
order. $K$ is presently known to the order
$\alpha_s^2$. After the Mellin transform (\ref{mellin}), the BFKL for the
gluon-gluon elastic forward scattering is
\begin{equation}
\label{nbfkl}
NF = \delta^{(2)}(k_1 -k_2) + (K_0 + K_1)F ,
\end{equation}
where $K_0$  is the leading order (LO) integral operator and $K_1$ is the
non-leading order (NLO) one. In particular,
(\ref{nbfkl}) is
\begin{equation}
\label{k}
K_0F =\frac{\alpha_sN_c}{\pi} \int\frac{d^2k}{(k_1 - k)^2}\left[F(N,k,k_2) -
\frac{k^2_1}{k^2 + (k - k_1)^2}F(N,k_1,k_2) \right]
\end{equation}
for forward scattering.
$k_{1,2}$ in Eq.~(\ref{nbfkl}) are transversal (two-dimensional) external momenta of
the upper and the lower ladder gluons respectively.\\
Along with their advantages,  both
the DGLAP and BFKL have certain disadvantages.
The DGLAP is practically free of disadvantages when it operates in the
region of
large $x$ but when it is extrapolated into the region of small $x$,
 contributions  $\sim\ln x^k$  systematically neglected  in the DGLAP
to all
orders in $\alpha_s$ must be incorporated. Without them, the standard
DGLAP prediction for the asymptotically small-$x$ behavior is

\begin{equation}
\label{dglapas}
 F \sim \exp\left[\frac{4n_c}{\pi b}\ln{1/x}
\ln\frac{\ln(Q^2/\Lambda^2_{QCD})}{\ln(\mu^2/\Lambda^2_{QCD})}\right]^{1/2}
\end{equation}
where $n_c = 3$,  $b = (11n_c - 2n_f)/12\pi$, $n_f$ is the number of involved
quark flavours and $\mu^2$ is a starting point for
$Q^2$ -evolution. \\
 The BFKL accounts for contributions $\sim\ln^k x$ to all
powers in the QCD coupling.
However from the theoretical point of view, the main disadvantage of the BFKL
is that it treats $\alpha_s$ as a constant.
Thus, its small-$x$ prediction
\begin{equation}
\label{bfklas}
F\sim x^{-\Delta_P}\sqrt{Q^2}
\end{equation}
with the Pomeron intercept
\begin{equation}
\label{ln2}
\Delta_P = \frac{\alpha_s N}{\pi} 4\ln2 \left[1 +
r \frac{\alpha_s}{\pi} \right]
\end{equation}
cannot be used without specifying a value of $\alpha_s$. In the
$\bar{M}\bar{S}$-scheme used in Eq.~(\ref{ln2}), $r =-20$ for four involved
quark flavors and $\alpha_s =0.2$. Thus,
the NLO-corrections (the second term in (\ref{ln2})) are enormously big.
Basically, recent works on the unpolarized DIS structure functions are
intendant
to overcome these disadvantages of the DGLAP and the BFKL
with different means. \\
The first possible solution to the problem of the big NLO corrections in
Eq.~(\ref{ln2})
is suggested in \cite{b}. In difference to the $\bar{M}\bar{S}$- scheme used for
obtaining
(\ref{ln2}), they suggest to use  the approach of
work~\cite{blm} on the optimal scale setting in (\ref{ln2}).
With it, the value of $\alpha_s$, still considered as
fixed, is decreased down to  $\alpha_s = 0.15$ and at the same time the
value of $r$ becomes lesser:
$\bar{r}\approx r/2$. Another improvement of the BFKL in  the framework of
this approach is that
the new scale does not lead to negative cross-sections for non-leading modes
as it was when $\bar{M}\bar{S}$ was used. \\
On the contrary, $\alpha_s$ is supposed to be really running in works
\cite{thorne}. The main assumption made there is that $\alpha_s$ in the
BFKL -kernel (\ref{k}) should depend on virtuality $k^2_1$  of the
external ladder gluon. The main
conclusion is that when  running $\alpha_s(k_1^2)$  is taken into
account, this new
version of the BFKL predicts a small-$x$ behavior
of the DGLAP type rather then
the power-like (the Regge-like) behaviour (\ref{bfklas}). \\
The QCD coupling $\alpha_s$ in the BFKL- kernel is supposed to be running
also  in works
\cite{ccs}, though its argument is more complicated
compared to
\cite{thorne}. Indeed,  Eq. (\ref{k}) reads that the integration
region over $k$ includes the
subregion~(i) where
\begin{equation}
\label{i}
k^2 <k^2_1
\end{equation}
 as well as the opposite subregion~(ii) where
\begin{equation}
\label{ii}
k^2>k_1^2 .
\end{equation}
At last, there is a subregion~(iii)
where
\begin{equation}
\label{iii}
k^2 \approx k^2_1.
\end{equation}
Applying  the Mellin transform (\ref{mellin}) to the BFKL and, after that,
applying to the result the new
Mellin transform
\begin{equation}
\label{mmellin}
F(N,t) = \int_{-\imath\infty}^{\imath\infty} \frac{dM}{2\pi\imath}
e^{Mt}F(M)
\end{equation}
converts the integral equation (\ref{bfkl})
into an algebraic one:
\begin{equation}
\label{mbfkl}
 NF = F_0 + \chi(M)F.
\end{equation}
$\chi(M)$ in (\ref{mbfkl}) is the characteristic function of the BFKL.
As the integral operator $K$ in
(\ref{bfkl}) consists of the leading order contribution $K_0$ and the
non-leading one $K_1$, $\chi = \chi_0 + \chi_1$ respectively. The LO
characteristic function
\begin{equation}
\label{xi0}
\chi_0 = \frac{\alpha_sNc}{\pi}\left[2\psi(1) -\psi(M) -\psi(1-M) \right] ,
\end{equation}
with $\psi$ being the Euler $\Gamma$ -function derivative.
Eq.~(\ref{xi0}) reads that $\chi_0$ is invariant
to replacing $M$ by $1-M$, also it has
singularities at $M=0$ and $M=1$. With good accuracy,
\begin{equation}
\label{xi0col}
\chi_0 \approx \chi_0^{col} = \frac{\alpha_s N_c}{\pi}
\left[\frac{1}{M} + \frac{1}{1-M} \right].
\end{equation}
Corespondingly, $K_0$ can be approximated by $K_0^{col}$ through the inverse
Mellin transform of (\ref{xi0col}). This approximation leaves out kinematical
region (\ref{iii}) but it is supposed in \cite{ccs} to be not important.
Suggesting that in  remaining regions (\ref{i}) and (\ref{ii})
$\alpha_s$ in $K_0^{col}$ should depend on
the largest gluon virtuality, works \cite{ccs} assume that
$\alpha_s$ depends on
$k_1^2$ in the region (\ref{i}) and on $k^2$ in the region
(\ref{ii}).  Then, using similar arguments, works \cite{ccs}
approximate the NLO BFKL characteristic function $\chi_1$ by
\begin{equation}
\label{xi1col}
\chi_1^{col} =(\alpha_s N_c/\pi)^2 (-11/12)
\left[\frac{1}{M^2} + \frac{1}{(1-M)^2} \right]
\end{equation}
so that discrepancy between $\chi_1$ and  $\chi_1^{col}$ is less than 7
per cents.
The number (-11/12) in (\ref{xi1col}) is
the subleading contribution to
the leading order DGLAP gluon anomalous dimension
$\gamma_{gg} = 1/\omega - 11/12$.
Approximations (\ref{xi0col}), (\ref{xi1col}) in works
\cite{ccs} and their treat of $\alpha_s$ are expected to imitate
next-to- next order contributions (and so on) to the BFKL in a rather
technically simple  way and still to be in a reasonable
accordance with the BFKL. \\
The approach using the BFKL for improving knowledge of the DGLAP anomalous
dimensions and splitting functions is suggested in works \cite{abf}.
The point is that, instead of straightforward procedure of accounting for
non leading order contributions to $\gamma_{gg}$ to all orders in $\alpha_s$,
one can use for it the BFKL which already includes such a resummation.
After the double Mellin transform both with respect to $x$ and with respect to
$\ln Q^2$ (we remind that we keep notations $\xi = \ln(1/x)$ and
$t = \ln{q^2/\mu^2}$ through this paper),

\begin{equation}
\label{dmellin}
F(\xi, t) = \int_{-\imath\infty}^{\imath\infty} \frac{dM}{2\pi\imath}
\int_{-\imath\infty}^{\imath\infty} \frac{dN}{2\pi\imath}F(M)
e^{Mt + N\xi}F(M,N)
\end{equation}
the BFKL is (cf. Eq.~(\ref{mbfkl}))

\begin{equation}
\label{dbfkl}
NF(M,N) = F_0 + \chi(M)F(M,N)
\end{equation}
with the obvious solution
\begin{equation}
\label{dm}
F(M,N) = \frac{F_0}{N -  \chi(M)} .
\end{equation}
Eq.~(\ref{dm}) reads that there is the relation between singularities of
$F$ in $M$ and in $N$. Indeed,
\begin{equation}
\label{dn}
N =  \chi(M) .
\end{equation}
at  points of singularity of Eq.~(\ref{dm}). \\
On the other hand, the DGLAP reads the asymptotically
small-$x$ behavior as  $F \sim\exp[\gamma(N) t]$  at such points,
which being compared to (\ref{dmellin}) immediately gives
\begin{equation}
\label{ddm}
M = \gamma(N) .
\end{equation}
Combining Eqs.~(\ref{dn}) and (\ref{ddm}) leads to
\begin{equation}
\label{d}
N = \chi(\gamma(N)), ~~~~M = \gamma(\chi(M))
\end{equation}
Obtained in \cite{j,bf} Eqs.~(\ref{d}) are called the
duality relations.  As $\chi$ contains contributions $\sim (\alpha_s/M)^k$
to all powers in $\alpha_s$
the relation $N = \chi(\gamma(N))$ can be used to express the anomalous
dimension $\gamma$ in terms of $\chi$ and its derivatives. However, it
cannot be done straightforwardly because $\chi$ has singularities
 at values of its argument $\gamma$ equal to 0 and 1. Due to
momentum conservation $\gamma(1) = 0$,  therefore it must be
\begin{equation}
\label{momcon}
\chi(\gamma(1))= \chi(0) = 1
\end{equation}
which contradicts to its actual behaviour
$\chi\sim\alpha_s/M$ (see (\ref{xi0})). This
contradiction can be corrected by
regularization:  $\chi\sim\alpha_s/M$ must be replaced at small $M$ by
\begin{equation}
\label{reg}
\chi\sim\alpha_s/(M + \alpha_s).
\end{equation}
Besides the regularization (\ref{reg}), $\chi(M)$ must be regularized at
$M \sim 1$ . It cannot be done with using $\chi(M)$ at $M \sim 1/2$ or so
because the solution of the BFKL is unstable at $M = 1/2$: the
NLO corrections (the second term in rhs of (\ref{ln2})) are quite
comparable to the leading order contribution.
 So, in \cite{abf} such regularization is done through introducing a new
parameter $\lambda$ which is a  "true"
value of the Pomeron intercept so that the new improved $\xi$  is
supposed to be stable at vicinity $M = \lambda$. Thus, in the context of
\cite{abf},
the expressions for new gluon anomalous dimensions contain a presently unknown
parameter $\lambda$
which must be fixed from experimental data. With such regularizations of
$\chi$, the new anomalous
dimensions incorporating NLO contributions to all powers in the QCD coupling
can be really expressed in terms of $\chi$ and its derivatives.\\
The next group of works deals with non-perturbative approaches to the Pomeron.
In \cite{kl}, the contribution of the four-gluon vertices to the Pomeron ladder
is considered. It is known that such a contribution is negligibly
small in the
perturbative QCD compared to the three-gluon vertices contribution.
However,  \cite{kl} notes that it is proportional to the trace of the
energy-momentum tensor in the chiral limit of massless quarks. Therefore,
the gluon ladder with the four-gluon vertices,
apart from a small perturbative contribution
\begin{equation}
\label{deltap}
\Delta = \frac{18\pi^2}{b^2}\int\frac{dM^2}{M^6} \rho^{pert}(M^2)
\end{equation}
to the Pomeron intercept $\Delta$
(in Eq.~(\ref{deltap}) $b = (33 -2n_f)/12\pi^2$),
may contain a non-perturbative contribution which is proportional to the
correlator \mbox{$<\theta_{\mu}^{\mu}(x)\theta_{\nu}^{\nu}(y)>$} of the
energy-momentum tensors.
Substituting the  spectral density $\rho$ of the correlator in the chiral
limit,
\begin{equation}
\label{rho}
\rho = (3/32\pi^2)M^4,
\end{equation}
into Eq.~(\ref{deltap}) leads to logarithmic dependency $\delta$ on the
upper limit of integration $M_0$ in (\ref{deltap}).  Estimating
\begin{equation}
\label{m0}
M^2_0 = 32\pi\sqrt{\epsilon_{vac}/(n_f^2 - 1)}
\end{equation}
where the value for the vacuum energy $\epsilon_{vac} = -(0.24GeV)^4$
is taken from the sum rule analysis, they obtain
\begin{equation}
\label{deltakl}
\Delta = 0.08.
\end{equation}
The modification of the above approach was made in \cite{s}. In essence, it
comes down to considering Eq.~(\ref{deltap}) in the same non-perturbative context
and suggesting to replace $M^6$ in (\ref{deltap}) by
$M^2(M^2 + k_t^2)^2$ where $k_t^2$ is an
infrared
cut-off in the transverse momenta space for the exchanged gluons in the
gluon ladder.
Treating this cut-off as a scale for applicability of perturbative
methods and putting to be equal to the inverse size of the instanton,
$k_t^2 = (0.6 GeV)^2$, work \cite{s} concludes that  the Pomeron intercept
\begin{equation}
\label{deltas}
\Delta = 0.005.
\end{equation}
Although both previous estimate of \cite{kl} and
this estimate are in a good agreement with
experimental data, both works give no prescription what to do with
the BFKL contribution to the Pomeron intercept, though it is greater
than the obtained non-perturbative contributions. As NLO corrections to the
LO BFKL decrease the value of the LO Pomeron intercept (see \ref{ln2}),
works \cite{kl},
\cite{s} suggest that
next corrections may decrease its value down to zero and only non-perturbative
contributions (\ref{deltakl}),(\ref{deltas}) would have non-zero
contributions. However, it
is not clear what is the accuracy of the predictions
(\ref{deltakl}),(\ref{deltas}) and what could be corrections to them. \\
The works I have
discussed contain different improvements of the DGLAP 
and the BFKL to
make them more consistent. However, what is really
needed for the region of small $x$ is a two-dimensional evolution 
equation combining
evolutions in $x$ and in $Q^2$ so that $\alpha_s$ were running.
Such an equation, the infrared
evolution equation (IREE), has been obtained recently
in work \cite{egt} for studying the small-$x$ behaviour of
the non-singlet contribution  $f^{NS}$ to the structure function $F_1$.
The basic idea for constructing such equations is similar to constructing
the renormalization group equations (RGE) but instead of evolution
with respect to the ultraviolet cut-off, it exploits the infrared cut-off
evolution:
when calculating a structure function,
let us introduce the infrared cut-off $\mu$
in the transverse momentum space for all integrations over virtual
particle momenta. With such a regularization, the structure function is $\mu$-
dependent. Instead of fixing $\mu$, one can evolute the structure function
with respect to it. The difference
between the IREE and the RGE is that physical quantities like
cross-sections do not depend on the
ultraviolet cut-off but they have to depend on the infrared one, so
within the IREE approach, results depend on
parameter $\mu$.
Work \cite{egt} predicts a scaling-like behavior
\begin{equation}
\label{fns}
f^{NS} \approx \left(\frac{1}{x}\sqrt{\frac{Q^2}{\mu^2}}\right)^a
\end{equation}
with the exponent
\begin{equation}
\label{a}
a = 0.37.
\end{equation}
This value was obtained when both the leading logarithmic (double-logarithmic)
and non-leading (single-logarithmic) contributions, including the running
$\alpha_s$ effects, were taken into consideration. Contrary to the
DGLAP prediction (\ref{dglapas}), Eq.~(\ref{fns}) predicts the power-like
small-$x$ behaviour for $f^{NS}$. However, similarly to the DGLAP,
 $a$  in Eq.~(\ref{fns}) depends on $n_f$, $\Lambda_{QCD}$ and on input 
 $\mu$.
The technical difference between the Pomeron and $f^{NS}$  is that the
Pomeron gluon ladder is replaced by the quark ladder. Such
quark ladder is a technically simpler object compared to the Pomeron.  Thus,
apart from physical implications of results obtained in \cite{egt},
one can regard them as
tests for checking some of the assumptions made in the works
I discussed above.
Earlier, $a$ was calculated in works \cite{ber} in the leading,
double-logarithmic (DL) approximation where $\alpha_s$ was fixed.
This DL result for the non-singlet structure function at small $x$ 
is analogous to the
leading order BFKL prediction for
the singlet structure function (the Pomeron). Accounting for the non-leading
contributions in \cite{egt} was done in a model-independent way.
On the other hand, it is analogous to modifications of the BFKL
made in works \cite{b}-\cite{ccs}.
Therefore, it is possible and interesting to compare them.
Having  done so, we conclude:\\
(i) It is impossible to imitate the running $\alpha_s$ effects by choosing a
reasonable scale for fixed $\alpha_s$ in expressions for intercept of
$f^{NS}$. More precisely, the value of the scale for $\alpha_s$ strongly 
changes when different kinds of non-leading contributions are accounted for. 
  So, concerning results of \cite{b} we think that the scale setting 
would have to be done again if a new portion of non-leading contributions 
to the BFKL were accounted for.  \\
(ii) Contrary to assumptions made in \cite{thorne, ccs},
results of \cite{egt}
read that virtualities of ladder partons can be  arguments of $\alpha_s$ in
evolution equations
only at $x\sim 1$. When $x$ is small, the argument of $\alpha_s$ is
more complicated.
 Accounting for non-leading contributions leads to changing the exponent
$a$ in (\ref{fns}) but it does not change the fact (obtained in \cite{ber} where
$\alpha_s$ was fixed) that small $x$-dependence of
$f^{NS}$ is power-like. Thus, we doubt the DGLAP-like $x$-dependence obtained
in \cite{thorne} for the Pomeron.\\
The small-$x$ behaviour (\ref{fns}) of $f^{NS}$  involves a
new mass scale $\mu$, with $\mu^2<<Q^2$.
 Before that,  one could think of a Regge-like small-$x$ behaviour
$f^{NS}\sim\left( s/Q^2\right)^a$ instead of (\ref{fns}).
This $\mu$-dependence is due to the
fact that the QCD perturbative methods
cannot be used in the region of too small momenta. We think, such a
dependence should exist for the Pomeron too.\\

I am grateful to M.Greco and S.I.Troyan for useful discussions.

\end{document}